# AN END-TO-END STUTTERING DETECTION METHOD BASED ON CONFORMER AND BILSTM

*Xiaokang Liu[1], Changqing Xu[1], Yudong Yang[1], Lan Wang[1], Nan Yan[1]*

[1]Shenzhen Institute of Advanced Technology, Chinese Academy of Sciences, Shenzhen, 518055

## ABSTRACT

Stuttering is a neurodevelopmental speech disorder characterized by common speech symptoms such as pauses, exclamations, repetition, and prolongation. Speech-language pathologists typically assess the type and severity of stuttering by observing these symptoms. Many effective end-to-end methods exist for stuttering detection, but a commonly overlooked challenge is the uncertain relationship between tasks involved in this process. Using a suitable multi-task strategy could improve stuttering detection performance. This paper presents a novel stuttering event detection model designed to help speech-language pathologists assess both the type and severity of stuttering. First, the Conformer model extracts acoustic features from stuttered speech, followed by a Long Short-Term Memory (LSTM) network to capture contextual information. Finally, we explore multi-task learning for stuttering and propose an effective multi-task strategy. Experimental results show that our model outperforms current state-of-the-art methods for stuttering detection. In the SLT 2024 Stuttering Speech Challenge based on the AS-70 dataset [1], our model improved the mean F1 score by 24.8% compared to the baseline method and achieved first place. On this basis, we conducted relevant extensive experiments on LSTM and multi-task learning strategies respectively. The results show that our proposed method improved the mean F1 score by 39.8% compared to the baseline method.

## 1. INTRODUCTION

Stuttering is a major speech disorder affecting approximately 70 million people worldwide, accounting for about 1% of the total population [2], [3]. For individuals who stutter, stuttering not only impacts their social functioning but also has negative effects on their mental health. With the advancement of communication technologies, these impacts have shown a tendency to intensify. This study aims to achieve automatic detection of stuttering, which has several important applications. For instance, automatic detection can help speech therapists reduce the workload associated with manual calculations, allowing them to more efficiently assess the severity of stuttering. Additionally, it can provide people who stutter (PWS) with immediate feedback on their speech fluency.

Despite the numerous potential applications of stuttering detection, related research remains limited from the perspectives of signal processing and machine learning. Stuttering is a neurodevelopmental speech disorder defined as abnormal, persistent interruptions to the forward flow of normal speech, typically manifested as prolongations, blocks, and repetitions of syllables, words, or phrases [4]. Speech-language pathologists assess the severity of stuttering by observing these core symptoms, typically evaluating both the acoustic properties of speech and speech fluency. In terms of acoustic properties, blocks and prolongations are common focal points. Blocks refer to involuntary pauses within the natural flow of speech, while prolongations pertain to the extended duration of any stuttering event. Regarding speech fluency, repetitions, interjections, and insertions are primary considerations. Repetitions can include the repetition of syllables, words, and phrases. Interjections are additional filler words such as exclamations and discourse markers [5].

However, due to the heterogeneity and overlap of stuttering behaviors, stuttering event detection (SED) remains challenging, especially when data is limited [5], [6], [7]. Existing SED models can be broadly categorized into two types: those based on speech signal feature analysis and those based on speech fluency analysis. In speech signal feature-based methods, researchers employ speech features combined with machine learning (ML) algorithms, such as Mel-frequency cepstral coefficients (MFCC), formants, pitch, zero-crossing rate (ZCR), and energy. These features are then used with algorithms like support vector machines (SVM) and k-nearest neighbors (k-NN) [8], [9], [10], [11], [12]. These traditional methods rely on both acoustic and non-parametric feature extraction techniques [13] and are relatively effective in detecting certain stuttering events, such as blocks and prolongations [14]. On the other hand, speech fluency analysis methods excel in detecting repetitions, insertions, and interjections. For example, automatic speech recognition (ASR) methods have been shown to outperform signal-based methods in detecting repetition events [7]. Additionally, context-aware approaches, such as bidirectional long short-term memory networks (BI-LSTM), convolutional LSTM (ConvLSTM), and Transformers, have also proven effective in stuttering event detection [7], [15], [16], [17].

It is important to highlight that current stuttering detection methods fail to differentiate between the various manifestations of stuttering, often conflating them. Each manifestation not only operates on distinct principles but also exhibits different performances in speech. Particularly in multitasking scenarios, where subtasks may be unrelated or even contradictory, the effectiveness of the overall results can be compromised. Therefore, this paper focuses not only on the signal and fluency characteristics of speech but also delves into the nuances of multitasking distinctions in stuttering. In this work, we present a novel method for stuttering event detection aimed at simplifying speech signal representation and capturing contextual relationships in stuttering patients' speech. The model first employs the Conformer to extract acoustic information from stuttering speech, followed by a Long Short-Term Memory (LSTM) network to capture contextual dependencies. Additionally, we investigate the effect of multi-tasking strategies on model performance and identify the optimal approach. Experimental results demonstrate that our method outperforms current state-of-the-art techniques for stuttering detection. On the AS-70 dataset, our proposed model achieves an average F1 score improvement of 39.8% over the baseline across five tasks.

## 2. RELATED WORK

Despite the numerous potential applications of stuttering detection, it has received relatively little attention. Stuttering is a neurodevelopmental speech disorder that not only affects the acoustic properties of speech but also impacts the overall fluency of speech. Therefore, existing research methods analyze stuttering from both the perspective of speech signal processing and speech fluency.

### 2.1 Speech acoustic feature analysis

Research has shown that stuttering affects different formant characteristics, such as transitions and fluctuations of formants [3]. Current stuttering detection methods primarily utilize spectral features, such as Mel-frequency cepstral coefficients (MFCC) and linear predictive cepstral coefficients (LPCC), or their variants, to capture formant-related information. Additionally, other spectral features such as pitch, zero-crossing rate, brightness, and spectral spread are also used. These features are commonly employed to build statistical models such as Hidden Markov Models (HMM), Support Vector Machines (SVM), and Gaussian Mixture Models (GMM).

For instance, Jhawar et al. [9] described a feature extraction method based on MFCC, which was further utilized to measure disfluency in speech and classified using K-Nearest Neighbors (K-NN). Chee et al. [10] proposed using LPCC and K-NN for feature extraction and classification of stuttered speech repetitions and prolongations. Ramteke et al. [11] employed MFCC, formants, and brightness for feature extraction, and Dynamic Time Warping (DTW) was used to distinguish between fluent speakers and individuals who stutter. Savin et al. [12] utilized MFCC, formants, pitch, zero-crossing rate (ZCR), and energy as feature extraction, focusing on detecting repetitions and prolongations using Artificial Neural Networks (ANN). Mahesha et al. [8] described an algorithm for automatic segmentation of stuttered speech, employing two feature extraction methods: short-term average amplitude and spectral spread, combined with Maximum Auto-Correlation Value (MACV) technique for fluency disorder classification. Mahesha et al. [13] proposed a stuttering disfluency classification technique based on Gaussian Mixture Models (GMM), using MFCC to estimate GMM parameters, and evaluated performance regarding repetitions, prolongations, and interjections. Ghonem et al. [18] described a method named I-Vector for automatic classification of stuttered speech. Wisniewski et al. [19] suggested using Hidden Markov Models (HMM) for automatic classification of stuttering, without focusing on specific disfluencies, aiming to identify any extraneous sounds in speech signals. Mahesha et al. [20] detailed a technique based on Support Vector Machines (SVM) and GMM for classifying disfluent speech.

### 2.2 Speech fluency analysis

The common symptoms of stuttering, particularly pauses and repetitions, significantly impact speech fluency. Therefore, in recent years, there has been a growing focus on approaches that address long-term speech fluency.

For instance, Kourkounakis et al. [16] developed a hybrid model based on RESNET and Bidirectional Long Short-Term Memory (BI-LSTM), where spectrograms were used as acoustic features to directly detect and classify stuttering events from signals. Lea et al. [7] proposed a ConvLSTM model architecture where Mel-filterbanks, phoneme probabilities, pitch, and pronunciation served as input features to the model. Czyzewski et al. [17] introduced an end-to-end model utilizing squeeze-and-excitation blocks followed by BILSTM and global attention mechanism. Jouaiti et al. [21] created a real-time model for stuttering detection and classification, employing BILSTM, MFCC, phoneme probabilities, and classes for training. Sheikh et al. proposed StutterNet [22], utilizing a delay neural network tailored to capturing contextual aspects of disfluent speech, trained on MFCC input features. On the other hand, automatic speech recognition (ASR) methods are considered superior to signal-based methods in detecting repetition events [7]. ASR methods employ various techniques including Hidden Markov Models (HMM), task-oriented language models, and pre-trained ASR models [23] [24], [25] [26] [27]. The wav2vec 2.0 model has demonstrated the best performance in speech-related tasks. As wav2vec 2.0's pre-training corpus primarily consists of fluent speech, Bayerl et al. [15] retrained the network using available fluency data with default configurations and achieved improved results.

## 3. DATA

In this paper, the proposed method was evaluated using the AS-70 dataset. The AS-70 dataset comprises recordings of adults who stutter (AWS) and native Mandarin speakers. A total of 70 native Mandarin-speaking AWS participated in the recording sessions, including 24 females, resulting in a gender ratio of 1.9:1. Each participant underwent an hour-long recording session, consisting of two parts: dialogues and reading speech commands. The annotation guidelines specify five types of stuttering, including:

[]: Word/phrase repetition, used to mark the entire repetition of a character or phrase.

/b: Block, indicating an audible or silent pause due to breath or stuttering.

/p: Prolongation, referring to elongated phonemes.

/r: Sound repetition, indicating repeated phonemes that do not constitute a complete character.

/i: Interjection, representing filler characters caused by stuttering, such as 'uh', 'um', or 'er'. It's worth noting that naturally occurring interjections that do not disrupt speech flow are excluded.

In the AS-70 dataset, the stuttering event proportions per minute are 15.58% for the dialogue task and 8.11% for the command task. From a classification perspective, the dataset exhibits a low proportion of stuttering events, indicating an imbalance. Additionally, as depicted in Figure 1, there is also an imbalance in the proportions of different types of stuttering within stuttering events. This imbalance poses a challenge for detection models.

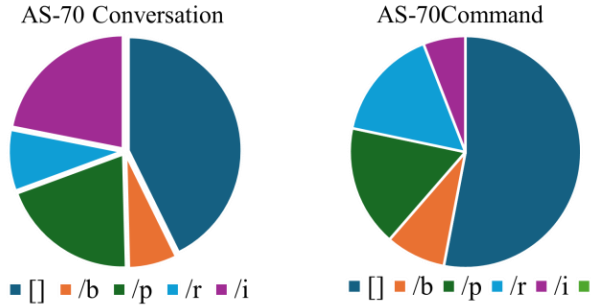

**Fig. 1** shows the proportion of symptoms in as-70 Conversation and AS-70 Command data respectively.

## 4. MODEL STRUCTURE

### 4.1. Baseline model

In this paper, we utilized the Conformer encoder [28] as the baseline model. The Conformer model consists of SpecAug, Convolution Subsampling, Linear, Dropout, and Conformer Blocks × N. Each Conformer Block comprises a Feed Forward Module, Multi-Head Self-Attention Module, Convolution Module, Feed Forward Module, and LayerNorm. Each module is followed by LayerNorm and Dropout, with residual connections where the residual data is the input data itself. The structure of the Conformer Encoder is depicted in Figure 2. In this study, we first extracted 80-dimensional fbank features from the speech, then fed them into the Conformer encoder, and finally performed classification through a linear layer, as illustrated in Figure 3(a).

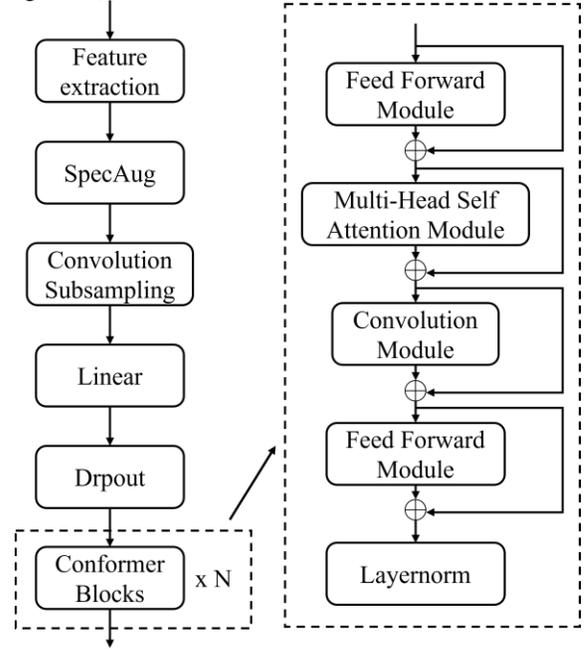

**Fig. 2.** Conformer structure

### 4.2 Proposed model

The symptoms of stuttering patients not only involve short-term speech information such as blocking and prolongation but also include long-term contextual issues such as semantic coherence and sequence in speech. Although using the Conformer model can effectively capture short-term speech information, it fails to consider the sequential and repetitive relationships of semantics in speech. In contrast, the BILSTM model can better capture longer-range bidirectional dependencies because it can learn which information to memorize and which to forget through the training process. Therefore, in this study, we introduced two layers of BILSTM after the baseline Conformer model to extract long-term contextual information in speech. The Conformer model consists of 6 layers, and a Conformer-ASR system trained on the AISHELL-1 dataset [29] was used as a pre-trained model. The number of nodes in these two layers of BILSTM is set to half the size of the Conformer output, i.e., 256, and the output size is half the input size, i.e., 128, as illustrated in Figure 3.

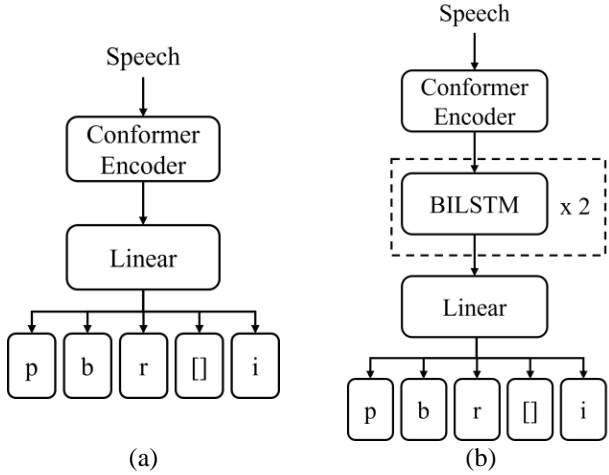

Fig. 3. Network structure of baseline model and Conformer-LSTM model. Where (a) is the baseline model and (b) is the proposed model.

### 4.3 Multi-Task Learning

During the training process of stuttering detection systems, a crucial issue is overfitting and lack of generalization [30]. Overfitting is particularly problematic in resource-limited scenarios, especially in detecting fluency disorders in stuttering therapy. To address this issue, multitask learning (MTL) is an effective regularization method. It encourages the model to generalize by introducing an additional auxiliary task that is different from but related to the primary task. Such models can benefit from regularization across tasks and auxiliary information. In fact, MTL has been successfully applied to various speech tasks, such as speech recognition or speech emotion recognition (SER).

Due to the high correlation between stuttering symptoms but the imbalance in their respective sample proportions, we chose to use other tasks as auxiliary tasks. We hypothesize that additional information and regularization effects can enhance the training effectiveness. To achieve this, we designed a model comprising five binary classification tasks, with the output layer defined as 2*5 dimensions, and used Binary CrossEntropy with Logits Loss (BCEWithLogitsLoss) as the loss function. For each sample $x_i$, its corresponding binary label is $y_i \in {0,1}$, and we want the model to predict the probability for each sample as $p_i \in [0,1]$. Therefore, BCEWithLogitsLoss can be expressed as follows:

$$loss = \frac{1}{N}\sum_{i=1}^{N}(y_i \cdot \log(\sigma(p_i)) + (i - y_i) \cdot \log(1 - \sigma(p_i))) \quad (1)$$

where $N$ is the number of samples, $\sigma(x) = \frac{1}{1 + e^{-x}}$. Is sigmoid function, and log is natural logarithm. This loss function measures the difference between the predicted probabilities and the actual binary labels, guiding the model to improve its predictions through the training process.

## 5. EXPERIMENTAL RESULTS

This section presents the experimental results of the proposed stuttering event detection model, which was trained and evaluated on the AS-70 dataset. We provide a detailed description of the experimental setup, evaluation metrics, and experimental outcomes. Additionally, we compare the performance of the proposed model with existing models and discuss the significance of the experimental results in the field of stuttering detection.

### 5.1. Setup

The model is built using the wenet library for feature extraction and audio processing. We used the Adam optimizer with a learning rate of 0.0001, combined with weighted entropy, focal loss, and an early stopping strategy with 10 patience epochs for model training. In this study, we followed the dataset split method used in the SLT2024 competition to divide the training, validation, and test sets, as shown in Table 1. This division ensures that speakers with different levels of stuttering severity are distributed across each partition.

Table 1. Division and Proportion of Dataset

| Dataset | number of people | number of files |
|---|---|---|
| Train | 43(61.42%) | 26659(64.09%) |
| Dev | 7(10%) | 4294(10.23%) |
| test | 20(28.57%) | 11000(26.21%) |

### 5.2. Evaluation

To rigorously evaluate the model's performance, we used the commonly employed metric in speech detection: Micro-F1. In the competition, we averaged the F1 scores of all tasks to obtain the final evaluation result, F1-final. The Micro-F1 score aggregates the contributions of all classes to compute the average metric, making it particularly suitable for evaluating the performance of models on datasets with imbalanced class distributions. The calculation formula is as follows:

$$F_1 = 2 \cdot \frac{precision \cdot recall}{precision + recall} \quad (2)$$

where Precision represents the ratio of the number of samples correctly predicted as positive by the model to the total number of samples predicted as positive. Recall represents the ratio of the number of samples correctly predicted as positive by the model to the total number of actual positive samples.

### 5.3. Results

To evaluate the performance of our model, we conducted five sets of experiments based on the AS-70 dataset. The first set of experiments explored the impact of the ASR pre-trained model on stuttering detection. The second set of experiments focused on an ablation study concerning the number of Conformer layers, aiming to determine the optimal number of layers for the stuttering task. The third set of experiments involved an ablation study of bidirectional long short-term memory networks (BILSTM), demonstrating the impact of different types of BILSTM on model performance. The fourth set of experiments compared the effectiveness of multi-task learning versus non-multi-task learning. The fifth set of experiments was used to compare the baseline results, the results we submitted in the competition, and the final results.

As shown in Table 2, the ASR pre-trained Conformer achieved an average F1 score that was 13.37% higher than the Conformer without pre-training, demonstrating greater robustness. This is particularly evident in task /b, where it had a significant impact. The Conformer used here has 3 layers, so the first row in Table 2 represents the locally reproduced baseline results. It should be noted that in the subsequent experiments, all Conformer models are pre-trained based on ASR.

**Table 2.** Comparison of the Conformer Model with and without ASR Transfer Learning

| ASR-pretrained | /p | /b | /r | [] | /i |
|---|---|---|---|---|---|
| No | 64.55 | 22.6 | 36.45 | 58.89 | 73.88 |
| **Yes** | **63.98** | **36.02** | **48.51** | **62.56** | **79.59** |

From Table 3, it can be observed that the Conformer12 model performs better on the five tasks: p, b, r, [], and i. In contrast, the Conformer3 model shows performance improvements of 6.12%, 5.96%, 26.18%, 17.88%, and 6.34% on these five tasks, respectively. Among the Conformer3, Conformer6, and Conformer12 models, performance gradually improves with an increase in the number of layers. However, in the Conformer15 model, the performance is lower than that of the Conformer12 model, indicating that increasing the number of layers leads to overfitting. Therefore, it can be concluded that the optimal number of layers for the Conformer model is 12.

**Table 3.** Comparison of Conformer Performance Across Various Layers

|  | /p | /b | /r | [] | /i |
|---|---|---|---|---|---|
| Conformer3 | 63.98 | 36.02 | 48.51 | 62.56 | 79.59 |
| Conformer6 | 67.71 | 42.9 | 55.75 | 66.56 | 82.03 |
| **Conformer12** | 67.9 | **38.17** | **61.21** | 73.75 | **84.64** |
| Conformer15 | **70.75** | 30.61 | 54.64 | **74.63** | 84.1 |

In the confusion experiments with LSTM, the short-term feature extractor was set to 12 layers of Conformer by default. As shown in Table 4, BILSTM outperforms LSTM, indicating that bidirectional dependency information is more effective for the stuttering task. Furthermore, Table 4 demonstrates that different stuttering tasks appear to be suited to different numbers of BILSTM layers. For example, the optimal number of BILSTM layers is two for tasks /p and /i, one for tasks /b and /r, and three for task []. This suggests that different tasks in stuttering are not closely related.

**Table 4.** Performance Comparison of BiLSTM Across Different Layers

| Model | /p | /b | /r | [] | /i |
|---|---|---|---|---|---|
| LISTM2 | 31.4 | 19.73 | 32.58 | 37.44 | 47.24 |
| BILSTM1 | 67.87 | **42.2** | **62.32** | 74.59 | 81.68 |
| **BILSTM2** | **68.12** | 36.69 | 60.01 | 73.62 | **85.11** |
| BILSTM3 | 61.95 | 35.77 | 60.48 | **75.21** | 80.5 |

In the multitask comparison experiments, 12 layers of Conformer and 2 layers of BILSTM were used as the base network. As shown in Table 5, multitask learning effectively addresses the overfitting problem of the proposed model but also limits the optimal performance for some tasks. When considering tasks individually, tasks /b and /r improve by 49.58% and 13.33%, respectively, compared to multitask learning. However, for tasks /p, [], and /i, performance decreases by 3%, 4%, and 18.5%, respectively. This indicates that tasks /p, [], and /i benefit from multitask learning. Furthermore, to verify the correlation between different stuttering tasks, a multitask experiment was designed by combining tasks /p, [], and /i. The results show that this three-task combination outperforms the multitask setup involving all five tasks, as shown in Table 6.

**Table 5.** Comparison of Multi-Task and Single-Task Performance

| MTL | /p | /b | /r | [] | /i |
|---|---|---|---|---|---|
| Yes | **68.59** | 36.98 | 60.01 | **73.62** | **85.11** |
| No | 66.55 | **73.35** | **69.24** | 70.75 | 29.77 |

**Table 6.** Comparison of Multi-Task Performance Across 3 and 5 Tasks

| MTL | /p | [] | /i | /b | /r |
|---|---|---|---|---|---|
| 5task | 68.59 | 73.62 | **85.11** | 36.98 | 60.01 |
| 3task | **70.89** | **76.06** | 84.64 | --- | --- |

Table 7 presents a comparison between the baseline model results, the results submitted during the competition, and the best results obtained through further exploration. The submitted results are a combination of the best outcomes from different models. Compared to the submitted results, the final results show significant improvements in tasks /b, /r, and [] by 66.43%, 15.82%, and 1.69% respectively. In the OUR* method, tasks /p and [] utilize tri-task multi-task training, task /i employs five-task multi-task training, while tasks /b and /r are trained separately.

**Table 7.** Comparison of Results Between the Proposed Method and Baseline Methods ("Ours" refers to the results

submitted in the competition, while "Ours*" denotes the results after further experimentation)

| Model | /p | /b | /r | [] | /i |
|---|---|---|---|---|---|
| Baseline | 65.12 | 24.3 | 41.86 | 61.85 | 74.87 |
| OURS | **70.89** | 44.07 | 59.78 | 74.79 | **85.15** |
| **OURS*** | **70.89** | **73.35** | **69.24** | 76.06 | **85.15** |

## 6. CONCLUSION AND FUTURE WORK

This paper introduces a novel stuttering event detection model that combines the Conformer model with a Long Short-Term Memory (LSTM) network to address key challenges in model generalization and data limitations in stuttering research. By using the Conformer model to extract acoustic features from stuttered speech, incorporating LSTM to capture contextual relationships, and implementing an optimized multi-task strategy, our model demonstrates strong performance in stuttering detection tasks. Experimental results indicate that the proposed model significantly outperforms current state-of-the-art methods, underscoring its effectiveness in improving both the accuracy and practicality of stuttering detection. This advancement offers speech-language pathologists a powerful tool for assessing both the type and severity of stuttering.

Future work will focus on analyzing the deeper relationships between tasks in stuttering detection to enhance interpretability from a theoretical perspective. Additionally, we will further optimize model performance by integrating acoustic signals with semantic information and explore the model's application and scalability across additional stuttering datasets.

## 7. ACKNOWLEDGMENTS

Do not include acknowledgments in the initial version of the paper submitted for blind review. If a paper is accepted, the final camera-ready version can (and probably should) include acknowledgments.